\documentclass{article}
  
\usepackage[utf8]{inputenc}
\usepackage[T1]{fontenc}
\usepackage[mathscr]{euscript}
\usepackage{geometry}
\usepackage{amssymb}
\usepackage{amsmath}
\title{A model of regular black hole satisfying the Weak Energy Condition}
\author{Stefano Chinaglia\footnote{e-mail: s.chinaglia@unitn.it} \\
\smallskip \\
\it Dipartimento di Fisica, Università di Trento \\
\it Via Sommarive 14, 38123 Trento, Italia \\
\smallskip \\
\it TIFPA (INFN) \\
\it Via Sommarive 14, 38123 Trento, Italia}
\date{\today}

\begin{document}
\maketitle

\begin{abstract}

In this article we discuss a star generated by some matter fluid, whose stress-energy tensor is known. We investigate both the maximally symmetric framework, where all the pressures are equal, and the dark energy framework, with $P = -\rho$. We show that the first one is not able to produce a black hole, while second one actually is. Finally, within this second framework, we present an example of regular black hole, showing it satisfies the NEC and the WEC.

\end{abstract}

\section{Introduction}

A typical subject in black hole literature is the search for regular black holes (RBH). As is well known, the first and most paradigmatic solution of the Einstein Equations (EE), the Schwarzschild solution \cite{Schwarzschild}, is singular at the origin. Other important solution, such as that of Reissner-Nordström (RN) \cite{Reissner} \cite{Nordstrom}, are also singular. However singularities are thought to be absent in nature, so that their presence indicates a breakout or an incompleteness of the theory.

The existence of black holes has been recently confirmed by the detection of gravitational waves by the LIGO-VIRGO collaboration \cite{LIGO}, meaning that the search for RBH has become more interesting. A first attempt to regularize the RN metric was done by Duan in 1954 \cite{Duan}, while the first example of RBH was proposed by Bardeen during the GR5 conference fourteen years later \cite{Bardeen} and since then many other examples have been provided. A non-exhaustive list of works on RBH includes \cite{Hayward_1}, \cite{Bronnikov}, \cite{Elizalde}, \cite{Dymnikova}, \cite{Dymnikova92}, \cite{NSS}, \cite{ANSS}, \cite{Modesto}, \cite{CMSZ}, \cite{Dymnikova_2}, \cite{Culetu}, \cite{Maeda}, \cite{Horava}, \cite{Horava_1}, \cite{KS}, \cite{CCO}, \cite{Pradhan}, \cite{Ma}, \cite{Johannsen}, \cite{Rodrigues}, \cite{Fan}, \cite{Beato}, \cite{Frolov_1}, \cite{Frolov_2}, \cite{Frolov_3}, \cite{Frolov_4}, \cite{Frolov_5}, \cite{Frolov_6}, \cite{Frolov_7}. For a review, see \cite{ans}.

Each of these has an asymptotic Schwarzschild behaviour, either when the observer is far from the source, or when the deformation parameters are set to 0, while at the centre they present a de Sitter. This is a necessary condition to regularize the metric, known as Sakharov criterion \cite{Sakharov}: namely the stress energy tensor reads $T_{\mu\nu} \simeq \Lambda_0 g_{\mu\nu}$ near the origin (as an example, see the behaviour of Hayward solution \cite{Hayward_1} as function of its single parameter $l$).

Among the various approaches to the problem, the most popular two involve either a modification of the stress-energy tensor, including some kind of exotic matter, or a modification of the Einstein tensor. It is worth to mention also an intermediate approach, which introduces a non trivial coupling between matter and geometry; examples are provided by \cite{BLZ} or \cite{Sert}.

In our work, we follow the first path, i.e. we take the geometric sector of the theory as in the standard GR, while we assume the stress-energy tensor being produced by some exotic fluid. However, contrary to what is typically done, we don't assume the source being point-like (as is done, for example, by Bardeen and Hayward and in non-linear electrodynamics): we consider as source an extended distribution, with a finite classical radius. It is worth to note that our approach differs also from that of \cite{NSS} and \cite{ANSS}: indeed they modify the stress-energy tensor, due to some quantum mechanism, in order to produce an effective mass distribution, which smears the delta of the Schwarzschild distribution; however, the classical limit is again a point-like mass generating the Schwarzschild black hole. We will show that our arrangement is crucial, in order to avoid singularities.

Finally, despite some similarities, our work is not related to \cite{Ovalle}, since it considers a gravitational source (which may also be a black hole) made of two different contributions and works out to separate their effects. Here we always deal with a single source, although we study both the isotropic and the anisotropic case. We also mention that Babichev {\it et al.} produced some papers concerning black holes in presence of exotic fluids (namely \cite{Babichev_1}, \cite{Babichev_2}, \cite{Babichev_2}), but their objectives and results are different from ours.

The paper is organized as follows: in Sec. 2 we give the general settings of the solution, while in Sec. 3 we study its properties in the most interesting regimes; in Sec. 4 we discuss the Buchdahl limit, showing that only suitable assumptions on the matter are capable to produce a black hole and to avoid the divergence of the central pressure; finally, in Sec. 5-7 we introduce and discuss an example solution and in Sec. 8 we have some concluding remark. Throughout the paper, unless otherwise specified, we use $c = \hbar = G = 1$.

\section{Basic assumptions}

We consider a spherically symmetric and static symmetry (SSSS), described by the metric

\begin{equation}
\label{metric}
ds^2 = - f(r) dt^2 + \frac{1}{g(r)} dr^2 + r^2 d\Omega^2 \ ,
\end{equation}

where $f$ and $g$ are suitable functions and $d\Omega^2$ is the volume element of a 2-sphere of unit radius. We also assume that this metric is generated by some fluid, whose stress-energy tensor is written in the form

\begin{equation}
\label{stress-energy}
T_\nu^\mu = \rho u^\mu u_\nu + (P - P_\bot) \delta_\nu^1 \delta_1^\mu + P_\bot \left( \delta_\nu^\mu + u^\mu\ u_\nu \right) \ ,\end{equation}

where $u_\mu$ is the four-velocity of the fluid, $\rho$ its density, $P$ the radial pressure and $P_\bot$ the transversal pressure. Concerning the density, we ask its positivity, $\rho \geq 0$; its monotonicity, $\rho' \leq 0$; and its finiteness, $\rho < \infty$.

In the system where the fluid is at rest, $u_\mu = (\sqrt{f}, 0, 0, 0)$ and the stress-energy tensor reads $T_\nu^\mu = {\rm diag}(\rho, P, P_\bot, P_\bot)$. This simplifies the EE, whose independent components become

\begin{equation}
\label{Einstein_1}
\frac{d}{dr} \left( r(1 - g) \right) = 8 \pi r^2 \rho \ ,
\end{equation}

\begin{equation}
\label{Einstein_2}
\frac{g}{fr} f' + \frac{g-1}{r^2} =8 \pi P \ .
\end{equation}

Moreover, we have to fulfill the conservation equation for the fluid $\nabla_\mu T_r^\mu = 0$, i.e.

\begin{equation}
\label{conservation}
P' +\frac{P+\rho}{2} \frac{f'}{f} + \frac{2 \left( P - P_\bot \right)}{r} = 0 \ .
\end{equation}

In order to further simplify the discussion, write $f = e^\alpha g$, where $\alpha$ is some arbitrary function. If so, eq. (\ref{Einstein_1})-(\ref{conservation}) read respectively

\begin{equation}
\label{field_1}
g(r) = 1 - \frac{c}{r} - \frac{8 \pi}{r} \int{r^2 \rho} \ ,
\end{equation}

\begin{equation}
\label{field_2}
8 \pi P = \frac{g \alpha'}{r} - 8 \pi \rho \ ,
\end{equation}

\begin{equation}
\label{field_3}
P' +\frac{P+\rho}{2} \left( \alpha' + \frac{g'}{g} \right) + \frac{2 \left( P - P_\bot \right)}{r} = 0 \ .
\end{equation}

where $c$ is an integration constant.

\section{Study of the solution}

There are two main possibilities, in order to solve system (\ref{field_1})-(\ref{field_3}): either we impose $P = P_\bot$, that is a symmetry condition on the generating fluid, or we fix $\alpha = 0$ (or any other constant as well), that is a condition on the resulting metric. The first choice, first performed by Schwarzschild in his original paper \cite{Schwarzschild}, is a very popular one, performed also by more recent works, such as \cite{Ovalle} (as starting point) and \cite{Mazur_Mottola}. Here we study separately the two cases, both in the classical regime, where we can distinguish  an inner and an outer part, and in some deformation regime.

\subsection{Case 1: $\alpha \neq 0$ and $P = P_\bot$}

Under these assumptions, eq. (\ref{field_1})-(\ref{field_3}) become

\begin{equation}
\label{field_1_1}
g(r) = 1 - \frac{c}{r} - \frac{8 \pi}{r} \int{r^2 \rho} \ ,
\end{equation}

\begin{equation}
\label{field_2_1}
\alpha' = 8 \pi \frac{r(P+\rho)}{1 - \frac{c}{r} - \frac{8 \pi}{r} \int{r^2 \rho}} \ ,
\end{equation}

\begin{equation}
\label{field_3_1}
P' + \frac{4 \pi}{1 - \frac{c}{r} - \frac{8 \pi}{r} \int{r^2 \rho}} \left( P^2 r + \left( \frac{c}{8 \pi r^2} + \rho r + \frac{1}{r^2} \int{r^2 \rho} \right) P + \frac{c}{8 \pi r^2} \rho + \frac{\rho}{r^2} \int{r^2 \rho} \right) = 0 \ .
\end{equation}

Once the density is given, eq. (\ref{field_1_1}) is solved autonomously, while eq. (\ref{field_2_1}) and (\ref{field_3_1}) form a closed system. However eq. (\ref{field_3_1}) cannot be solved analytically, unless trivial cases.

In order to fix ideas, study the classical limit of a finite star of radius $R$, where $\rho = \rho_0 \theta (R-r)$. If so, the three equations gain the simpler form

\begin{equation}
\label{field_1_1_const}
g(r) = 1 - \frac{8 \pi}{3} \rho_0 r^2 \ ,
\end{equation}

\begin{equation}
\label{field_2_1_const}
\alpha' = \frac{r(P+\rho)}{1 + \frac{c}{3} - \frac{8 \pi}{3} \rho_0 r^2} \ ,
\end{equation}

\begin{equation}
\label{field_3_1_const}
P' + \frac{4 \pi}{1 - \frac{c}{r} - \frac{8 \pi}{3} \rho_0 r^2} \left( P^2 r + \left( \frac{c}{8 \pi r^2} + \rho r + \frac{1}{3} \rho_0 r \right) P + \frac{c}{8 \pi r^2} \rho + \frac{\rho}{3} \rho_0 r \right) = 0 \ ,
\end{equation}

in the interior of the star; for its exterior, we have just

\begin{equation}
\label{field_1_1_const_ext}
g(r) = 1 + \frac{c}{r} - \frac{2M}{r} \ ,
\end{equation}

\begin{equation}
\label{field_2_1_const_ext}
\alpha' = 0 \ ,
\end{equation}

\begin{equation}
\label{field_3_1_const_ext}
P = 0 \ .
\end{equation}

For the first equation, $M$ is the whole mass of the star: $M = 4 \pi \int{r^2 \rho}$, where the integral is performed on the whole mass distribution (which is finite).

Eq. (\ref{field_1_1_const}) and eq. (\ref{field_1_1_const_ext}) impose to set $c=0$. Eq. (\ref{field_1_1_const}) represents the inner solution of a star and it contains only a de Sitter core; so

\begin{equation}
\label{radius_1}
g_{\rm int}(R) = 1 - \frac{8 \pi}{3} \rho_0 R^2 \ .
\end{equation}

On the other hand, calculating the solution on $R$ from outside, we have

\begin{equation}
\label{radius_2}
g_{\rm ext}(R) = 1 + \frac{c}{r} - \frac{2M}{R} = 1 +\frac{c}{r} - \frac{8 \pi}{3} \rho_0 R^2 \ ,
\end{equation}

and since the metric solution must be continuous, this imposes $c=0$ also outside. Moreover, as an extra proof, the external solution must be {\it exactly} Schwarzschild: and since eq. (\ref{field_1_1_const_ext}) already contains what is physically interpreted as the whole mass, $M$, $c$ would be an extra term and should be set to 0.

Nothing changes, using a more general density profile, since the $c/r$ term is the solution of the homogeneous equation $G_0^0 = 0$. Within our framework, the density profile modifies only the rhs of the equation and the general solution is of the form

\begin{equation}
\label{gen_sol}
g(r) = 1 + \frac{c}{r} + g_1 (r) \ ,
\end{equation}

where $g_1$ only depends on the choice of the density. This cannot be chosen so that $c/r$ cancels, since $c$ is not a free but arbitrary parameter, but an integration constant. However, every solution should reduce to the classical one and since the only deviation from classicality lays in $g_1$, we have that $g_1 \rightarrow \frac{8 \pi}{3} \rho_0 r^2$ for $r<R$ in the classical regime and the solution reduces to (\ref{field_1_1_const}): but no $c/r$ must appear inside the star, so the term vanishes and all the arguments used to set $c=0$ in the classical regime hold also for any other density.

This discussion solves the question arosed by P. Nicolini, A. Smailagic and E. Spallucci in their note \cite{Nicolini}. In our previous paper \cite{CZ}, we argued that the Schwarzschild term $c/r$ cannot be eliminated using the reconstruction approach (a very popoular one, carried e.g. by \cite{Dymnikova} or \cite{Beato}). Indeed, the presence of an integration constant $c$ is a mathematical fact that cannot be avoided; what can be done is to find a physical reason to set it to 0. Here we showed a powerful argument, but in general things do not behave so well.

$\rho=constant$ is one of the trivial cases in which the pressure can be found explicitly: indeed we solve eq. (\ref{field_3_1_const}) via separation of variables; we get

\begin{equation}
\label{pressure}
P(r) = \rho_0 \frac{ \sqrt{1 - \frac{8 \pi}{3} \rho_0 r^2} - \sqrt{1 - \frac{8 \pi}{3} \rho_0 R^2} }{ 3 \sqrt{1 - \frac{8 \pi}{3} \rho_0 R^2} - \sqrt{1 - \frac{8 \pi}{3} \rho_0 r^2} } \ .
\end{equation}

Coupling this with eq. (\ref{field_2_1_const}) we find also $\alpha (r)$:

\begin{equation}
\label{alpha}
\alpha (r) = 2 \ln{\left( \frac{ 3 \sqrt{1 - \frac{8 \pi}{3} \rho_0 R^2} - \sqrt{1 - \frac{8 \pi}{3} \rho_0 r^2} }{ 2 \sqrt{1 - \frac{8 \pi}{3} \rho_0 r^2} } \right) } \ ,
\end{equation}

so that the $g_{00}$ component of the internal solution is

\begin{equation}
\label{sol_int_const}
f(r) = \frac{1}{4} \left( 3 \sqrt{1 - \frac{8 \pi}{3} \rho_0 R^2} - \sqrt{1 - \frac{8 \pi}{3} \rho_0 r^2} \right)^2 \ ,
\end{equation}

while the external part is just the Schwarzschild term. Unfortunately, these simple results hold only in the case of $\rho = \rho_0$: indeed, it is easy to prove that otherwise eq. (\ref{field_3_2}) has no more separable variables. However, fortunately, this approach is not the best to produce RBH from some kind of gravitating matter.

\subsection{Case 2: $\alpha = 0$}

In this case, eq. (\ref{field_1})-(\ref{field_3}) reduce to

\begin{equation}
\label{field_1_2}
g(r) = 1 - \frac{c}{r} - \frac{8 \pi}{r} \int{r^2 \rho} \ ,
\end{equation}

\begin{equation}
\label{field_2_2}
P = - \rho \ ,
\end{equation}

\begin{equation}
\label{field_3_2}
P' + \frac{2\left( P - P_\bot \right)}{r} = 0 \ ,
\end{equation}

where $c$ is again an integration constant. We see that $c=0$ with the same arguments of Case 1, so we won't show it again.

Once the density $\rho$ is known, the only thing we are left with is calculating $P_\bot$:

\begin{equation}
\label{P_bot}
P_\bot = - \left( \rho + \frac{1}{2} \rho' r \right) \ .
\end{equation}

The sign of $P_\bot$ is not a priori determined, since $\rho$ and $\rho'$ have opposite signs.

Finally, notice that condition $\alpha = 0$ coincides with $P = - \rho$; this is more interesting, since $P = - \rho$ is a condition over the generating matter (and so it is a priori), instead of the condition on the solution $\alpha = 0$ (which is a posteriori).

To fix ideas, work once again in the simple case of a finite star with radius $R$ and constant density $\rho(r) = \rho_0 \theta (R-r)$. For $r>R$ equations easily read

\begin{equation}
\label{field_1_2_const}
g(r) = 1 - \frac{2M}{r} \ ,
\end{equation}

\begin{equation}
\label{field_2_2_const}
P = 0 \ ,
\end{equation}

\begin{equation}
\label{field_3_2_const}
P_\bot = 0 \ ,
\end{equation}

where $M \equiv 4 \pi \int{r^2 \rho}$ is the total mass of the star. On the other hand, for $r<R$, we have

\begin{equation}
\label{field_1_2_const_int}
g(r) = 1 - \frac{8 \pi}{3} \rho_0 r^2 \ ,
\end{equation}

\begin{equation}
\label{field_2_2_const_int}
P = - \rho_0 \ ,
\end{equation}

\begin{equation}
\label{field_3_2_const_int}
P_\bot = P \ .
\end{equation}

Outside the black hole there is no difference among different models. Notice also that $\rho = constant$ implies $ P = P_\bot = constant$, which forbids the pressure to vanish at the border of the star ($P(R) = 0$). One can object that $\rho = constant$ is just a toy model and with a more realistic matter distribution, such as

\begin{equation}
\label{toy_density_1}
\rho (r) = \rho_0 \left( 1 - \frac{r^2}{R^2} \right) \theta(R-r) \ ,
\end{equation}

then $P(R)=0$. On the other hand, this particular model is not quite good to be deformed, since it can produce negative densities. A more realistic model is given by

\begin{equation}
\label{toy_density_2}
\rho (r) = \rho_0 \frac{1 + e^{-R/\xi}}{1 + e^{(r-R)/\xi}} \ .
\end{equation}

In this model, the Heaviside function has been already deformed in the style of a Fermi-Dirac distribution, where parameter $\xi$ plays the role of some fundamental length. Later on, we will show that a density similar to (\ref{toy_density_2}) is actually able to produce a RBH.

\section{Buchdahl limit}

In this section, we show that the maximally symmetric $P = P_\bot$ case is not able to solve the so called Buchdahl limit \cite{Buchdahl}, i.e. the divergence of the central pressure; moreover, when we set $R$ lower than the Schwarzschild radius, the central pressure acquires an imaginary part.

\subsection{Buchdahl limit "from above"}

We firstly see what happens for $P = P_\bot$. Since eq. (\ref{field_3_1}) cannot easily be solved exactly, we take an approximate solution: around the origin, we write the density as

\begin{equation}
\label{central_density}
\rho(r \rightarrow 0) = \rho_0 - \kappa r^n + ... \ ,
\end{equation}

where $\rho_0$ is the central density, $\kappa > 0$ is some parameter and $n > 0$. Eq. (\ref{central_density}) directly descends from our requests about the density. Thus said, eq. (\ref{field_3_1}) reads

\begin{equation}
\label{field_3_1_approx_1}
\frac{2P'}{r} + 8 \pi \frac{ P^2 + \left( \frac{4}{3} \rho_0 - \frac{n+4}{n+3} \kappa r^n \right) P + (\rho_0 - \kappa r^n) \left( \frac{1}{3} \rho_0 - \frac{\kappa}{n+3} r^n \right) }{ 1 - \frac{8 \pi}{3} \rho_0 r^2 + \frac{8 \pi}{n+3} \kappa r^{n+2} } = 0 \ .
\end{equation}

The only way to hope in an analytical solution is to further approximate this object taking the lowest orders in $r$. The resulting equation is

\begin{equation}
\label{field_3_1_approx_2}
\frac{2P'}{ P^2 + \frac{4}{3} \rho_0 P + \frac{1}{3} \rho_0^2 } + \frac{8 \pi r}{1 - \frac{8 \pi }{3} \rho_0 r^2} \left[ 1 - \frac{1}{n+3} \left[ \frac{ (n+4) P + \frac{n+6}{3} \rho_0 }{ P^2 + \frac{4}{3} \rho_0 P + \frac{1}{3} \rho_0^2 } + \frac{8 \pi r^2}{1 - \frac{8 \pi}{3} \rho_0 r^2} \right] \kappa r^n \right] = 0 \ .
\end{equation}

But since we are working in the limit $r \rightarrow 0$, the pressure appearing in the square parentheses is just the central pressure $P_0$: any other correction is suppressed at the lowest order. This allows us to integrate both sides of eq. (\ref{field_3_1_approx_2}) in order to get the solution:

\begin{equation}
\label{central_pressure}
P(r \rightarrow 0) = \rho_0 \frac{ e^{8 \pi \kappa \rho_0 A r^{n+2} } \sqrt{1 - \frac{8 \pi}{3} \rho_0 r^2 } - \sqrt{ 1 - \frac{8 \pi}{3} \rho_0 R^2 } }{ 3 \sqrt{1 - \frac{8 \pi}{3} \rho_0 R^2 } - e^{8 \pi \kappa \rho_0 A r^{n+2} } \sqrt{1 - \frac{8 \pi}{3} \rho_0 r^2 } } \ ,
\end{equation}

where $A$ is the constant

\begin{equation}
\label{A}
A \equiv \frac{ (n+4) P_0 + \frac{n+6}{3} \rho_0 }{ 3(n+2)(n+3) \left( \left( P_0 + \frac{4}{3} \rho_0 \right) P_0 + \frac{1}{3} \rho_0^2 \right) } \ .
\end{equation}

Eq. (\ref{central_pressure}) allows to calculate the central pressure for a generic density profile; and surprisingly we have

\begin{equation}
\label{true_central_pressure}
P(r=0) = \rho_0 \frac{ 1 - \sqrt{ 1 - \frac{8 \pi}{3} \rho_0 R^2 } }{ 3 \sqrt{1 - \frac{8 \pi}{3} \rho_0 R^2 } - 1 } \ ,
\end{equation}

which is exactly the same of the classical case, taking eq. (\ref{pressure}) for $r=0$. It may sound strange, but it is just a natural consequence of the request $\rho (r \rightarrow 0) = \rho_0 - \kappa r^n + ...$, where the dominant term is the constant density of the classical limit. This means that, under the symmetry assumption $P = P_\bot$ there is no way to avoid the Buchdahl limit and pressure becomes infinite at $R = \frac{9}{8} R_S$, independently on the density profile of the star.

Notice that we implicitly assumed that $8 \pi \kappa \rho_0 r^{n+2} A$ in the limit $P_0 \rightarrow \infty$ remains finite. However it is not difficult to see that, in this limit,

\begin{equation}
\label{A_P_inf}
A(P_0 \rightarrow \infty) = \frac{n+4}{3(n+2)(n+3)P_0} \rightarrow 0 \ .
\end{equation}

Things look quite different, working with $\alpha =0$. In this case we have $P = - \rho$ and $P_\bot = P + \frac{1}{2} P'r$: so, a suitable choice of the density profile prevents any risk of divergence and fully avoids the Buchdahl limit.

\subsection{Buchdahl limit "from below"}

Before moving to the next section, see what happens when $R < \frac{9}{8} R_S$. In the case of $\alpha=0$, there is no risk, since the central pressure is fully determined (only) by the choice of the density. On the other hand, for the symmetry assumption $P = P_\bot$ things are different. In order to see it, rewrite eq. (\ref{true_central_pressure}) in terms of $R_S$: we have

\begin{equation}
\label{true_central_pressure_A}
P_0 = \frac{3R_S}{R^3} \frac{ 1 - \sqrt{1 - \frac{R_S}{R}} }{ 3 \sqrt{1 - \frac{R_S}{R}} - 1 } \ .
\end{equation}

This shows that the inner "Buchdahl zone" is divided in two regions: the first, with $R_S < R < \frac{9}{8} R_S$, where the central pressure becomes negative; and the second, for $R < R_S$, where it acquires an imaginary part.

In the first region, the central pressure varies from $-\frac{3R_S}{R^3}$ at $R_S$ to $-\infty$ at the Buchdahl point ($R = \frac{9}{8} R_S$). In the second, the central pressure reads

\begin{equation}
\label{true_central_pressure_B}
P_0 (R <R_S) = - \frac{\rho_0}{2 \left( \frac{9 R_S}{8R} - 1 \right)} \left( 1 - \frac{3 R_S}{4R} \pm \frac{i}{2} \left| {\sqrt{1 - \frac{R_S}{R}}} \right| \right) \ .
\end{equation}

Thus the imaginary part cannot be set to 0. But since the pressure must always be real, this means that the maximally symmetric fluid cannot build a star, whose radius is lower than the Schwarzschild radius $-$ i.e. a black hole. Moreover, since a star typically starts collapse when $R \gg R_S$, we expect that also the Buchdahl radius is never reached by the collapsing matter.

\section{A model of RBH satisfying the WEC}

We excluded the maximal symmetry assumption ($P = P_\bot$) as a source of generating black holes, so focus on the matter assumption ($P = - \rho$). Then we have to deal with eq. (\ref{field_1_2})-(\ref{field_3_2}). We choose the density profile so that it reduces to the classical density $\rho (r) = \rho_0 \theta (R-r)$ and prevents any divergence. A good choice could be (\ref{toy_density_2}), but a smarter one is

\begin{equation}
\label{smart_density}
\rho (r) = \rho_0 \frac{1+e^{-\lambda R^3}}{1 + e^{\lambda \left( r^3 - R^3 \right)}} \ ,
\end{equation}

where parameter $\lambda$ plays the role of quantum-like deformation; in the limit $\lambda \rightarrow \infty$, density (\ref{smart_density}) reduces to the classical one, as required; the use of cubic powers in the exponentials is chosen just for calculative convenience.

With this choice, the metric becomes

\begin{equation}
\label{smart_metric}
\begin{split}
g(r) &= 1 - \frac{8 \pi}{r} \rho_0 \left( 1 + e^{-\lambda R^3} \right) \int{ \frac{r^2}{1 + e^{\lambda \left( r^3 - R^3 \right)}} dr} \\
&= 1 - \frac{8 \pi \rho_0}{3r} \left( 1 + e^{-\lambda R^3} \right) \left( r^3 - \frac{1}{\lambda} \ln{\left( \frac{1 + e^{\lambda \left( r^3 - R^3 \right)}}{1 + e^{-\lambda R^3}} \right)} \right) \ ,
\end{split}
\end{equation}

which reduces to the classical solution (Schwarzschild outside + de Sitter inside) when $\lambda \rightarrow \infty$:

\begin{equation}
\label{limit_1}
g(r>R; \lambda \rightarrow \infty) = 1 - \frac{2M}{r} + o(\lambda^{-1}) \ ,
\end{equation}

\begin{equation}
\label{limit_2}
g(r<R; \lambda \rightarrow \infty) = 1 - \frac{8 \pi \rho_0}{3} r^2 + o(\lambda^{-1}) \ .
\end{equation}

\subsection{General properties of the horizon}

The first question in front of eq. (\ref{smart_metric}) is whether it represents or not a black hole. Unfortunately we are not able to find the horizon(s) explicitly, because the required equation $g(r_H) = 0$ is too difficult to be solved analytically. However we are still able to prove that, for suitable values of the central density $\rho_0$, horizons exist.

In order to show it, calculate the stationary points of the metric: equation $g'(r_{min})=0$ reads

\begin{equation}
\label{minimum}
\frac{1}{\lambda} \ln{ \left( \frac{1+e^{\lambda \left( r^3_{min} - R^3 \right)}}{1+e^{- \lambda R^3}} \right) } = r_{min}^3 \frac{e^{\lambda \left( r^3_{min} - R^3 \right)} - 2}{1+e^{\lambda \left( r^3_{min} - R^3 \right)}} \ .
\end{equation}

Of course there is no hope to solve it, but we can use it into eq. (\ref{smart_metric}) to find that

\begin{equation}
\label{minimum_A}
g(r_{min}) = 1 - 8 \pi \rho_0 \left(1 + e^{-\lambda R^3} \right) \frac{r^2_{min}}{1+e^{\lambda \left( r^3_{min} - R^3 \right)}} \ ,
\end{equation}

and since $\left(1 + e^{-\lambda R^3} \right) \frac{r^2_{min}}{1+e^{\lambda \left( r^3_{min} - R^3 \right)}}$ is always positive and the position of the minimum does not depend on $\rho_0$, we only need to tune $\rho_0$ in order to have $g(r_{min}) < 0$. We cannot proceed analytically, but we are at least sure that the minimum may be negative, so that there should be an horizon at some point.

After seeing that an horizon exists, we note that there is a minimal value for it: indeed, the lhs of eq. (\ref{minimum}) is strictly positive, so the rhs must be positive too; and this means that $r_{min} > \left( R^3 + \frac{1}{\lambda} \ln 2 \right)^{1/3}$. However, since the external horizon is more advanced than any negative minimum of $g(r)$, it follows that

\begin{equation}
\label{condition_1}
r_H > r_{min} > \left( R^3 + \frac{1}{\lambda} \ln 2 \right)^{1/3} \ .
\end{equation}

In the limit of $R \rightarrow 0$, classically corresponding to a point-like source, this reduces to $r_H > \left( \frac{1}{\lambda} \ln 2 \right)^{1/3}$, which correctly reduces to the classical limit $r_H > 0$ when $\lambda \rightarrow \infty$. This result is a strong clue of the quantum nature of $\lambda$: the impression is that $(\ln 2 / \lambda)^{1/3}$ is some kind of fundamental length and smaller lengths are not allowed.

\section{Energy conditions}

We anticipated that black hole metric (\ref{smart_metric}) fulfils the WEC; here is the time to prove it. In order to have a complete discussion, we also see what happens in the case of the Null Energy Condition (NEC), the Dominant Energy Condition (DEC) and the Strong Energy Condition (SEC).

The {\bf NEC} only consists in the positivity of the scalar $T_\mu^\nu k_\nu k^\mu$, where $k_\mu$ is a general null vector. Using our stress-energy tensor within the $P = - \rho$ and SSSS framework, we have

\begin{equation}
\label{NEC}
\begin{split}
T_\mu^\nu k_\nu k^\mu &= - (\rho + P_\bot) (k_0 k^0 + k_1 k^1) \\
&= \frac{1}{r^2} (\rho + P_\bot) \left( k_2^2 + \frac{k_3^2}{\sin^2 \theta} \right) \\
&\geq 0 \ ,
\end{split}
\end{equation}

meaning that $\rho + P_\bot \geq 0$, i.e.

\begin{equation}
\label{NEC_2}
\rho - \left( \rho + \frac{1}{2} \rho' r \right) = - \frac{1}{2} \rho' r \geq 0 \ .
\end{equation}

and this is easily verified, since $\rho' \leq 0$ was one of our original requests on the density.

Discuss now the {\bf WEC}. Always referring to our stress-energy tensor (\ref{stress-energy}) and considering the timelike vector $X_\mu$, it reads

\begin{equation}
\label{WEC}
\begin{split}
T_\mu^\nu X_\nu X^\mu &= - (\rho + P_\bot) (X_0 X^0 + X_1 X^1) - P_\bot \left| X_\mu X^\mu \right| \\
&= \frac{1}{r^2} (\rho + P_\bot) \left( X_2^2 + \frac{X_3^2}{\sin^2 \theta} \right) + \rho \left| X_\mu X^\mu \right| \\
&\geq 0 \ ,
\end{split}
\end{equation}

consisting in the two conditions

\begin{equation}
\label{WEC_1}
\rho \geq 0 \ ,
\end{equation}

\begin{equation}
\label{WEC_2}
\rho + \left( - \rho - \frac{1}{2} \rho' r \right) \geq 0 \ .
\end{equation}

The first one is easily satisfied, since is just one of our requests on the density. The second one holds from the discussion of the NEC.

For the {\bf DEC} we should prove, together with the WEC, that the vector $- T_\beta^\mu Y^\beta$ is causal future-directed, where $Y^\beta$ is any causal and future-directed vector. We already showed that the WEC holds, so we only need to check if

\begin{equation}
\label{DEC_0}
- T_\beta^0 Y^\beta \geq 0 \ ,
\end{equation}

\begin{equation}
\label{DEC_1}
\begin{split}
g_{\mu \nu} (- T_\beta^\mu Y^\beta) (- T_\alpha^\nu Y^\alpha) &= \rho^2 Y_0 Y^0 + P^2 Y_1 Y^1 + P_\bot (Y_2 Y^2 + Y_3 Y^3) \\
&= - \left( \frac{1}{r^2} (\rho^2 - P_\bot^2) \left( (Y_2)^2 + \frac{(Y_3)^2}{\sin^2 \theta} \right) + \rho^2 \left| Y_\beta Y^\beta \right| \right) \\
&\leq 0 \ .
\end{split}
\end{equation}

The first equation easily reads $\rho Y^0 \geq 0$ and since $\rho$ and $Y^0$ are both positive, it is immediately satisfied. Regarding the second, it is not difficult to see that this corresponds to the condition $\rho^2 - P_\bot^2 \geq 0$, i.e., since we know the relation between $\rho$ and $P_\bot$,

\begin{equation}
\label{DEC_2}
\rho \geq - \frac{1}{4} \rho' r \ .
\end{equation}

In general it is not possible to establish if this holds or not, since it depends on how $\rho'r$ is faster than $\rho$. We can however discuss it in the case of our density (\ref{smart_density}): we get

\begin{equation}
\label{DEC_3}
1 + \left( 1 - \frac{3}{4} \lambda r^3 \right) e^{\lambda (r^3 - R^3)} \geq 0 \ ,
\end{equation}

and it is not difficult to see that, for $r \rightarrow \infty$, this is badly violated. So, contrary to the NEC and the WEC, the DEC is violated.

The only thing we are left with is seeing whether the {\bf SEC} is satisfied or not. In other words, we need to see if $\left( T_\mu^\nu - \frac{1}{2} \delta_\mu^\nu \right) X_\nu X^\mu \geq 0$, where $T$ is the trace of the stress-energy tensor and $X_\mu$ any causal vector. We have

\begin{equation}
\label{SEC_1}
\begin{split}
\left( T_\mu^\nu - \frac{1}{2} \delta_\mu^\nu \right) X_\nu X^\mu &= - (\rho + P_\bot) (X_0 X^0 + X_1 X^1 ) - \frac{1}{2} (\rho - P) \left| X_\mu X^\mu \right| \\
&= (\rho + P_\bot) \left( (X_2)^2 + \frac{(X_3)^2}{\sin^2 \theta} \right) + P_\bot \left| X_\mu X^\mu \right| \\
&\geq 0 \ .
\end{split}
\end{equation}

This imposes the two conditions $\rho + P_\bot \geq 0$ and $P_\bot \geq 0$, which respectively read

\begin{equation}
\label{SEC_2}
- \frac{1}{2} \rho' r \geq 0
\end{equation}

\begin{equation}
\label{SEC_3}
\rho + \frac{1}{2} \rho' r \leq 0
\end{equation}

The first one is satisfied due to the NEC. On the other hand, we cannot discuss the second immediately, since $\rho$ and $\rho'$ have opposite signs and may have very different forms. However, if we want to verify that the SEC is violated, it is enough to show it is for some region of space. In order to do this, focus around the origin: in this limit, the density reads $\rho (r \rightarrow 0) = \rho_0 - \kappa r^n + ...$ with $\kappa > 0$, so that cond. (\ref{SEC_3}) reads

\begin{equation}
\label{SEC_violation}
\kappa \geq \frac{2 \rho_0}{(n+2) r^n} \ .
\end{equation}

But this should hold for any $r$ around the origin and it's clear it cannot be: the SEC is violated.

\section{Limit $R \rightarrow 0$}

As a last point, we see what happens in the limit $ R \rightarrow 0$. We already introduced this limit when speaking about the minimal value of the horizon, now it is time to investigate it better.

First of all, back to eq. (\ref{smart_metric}), the limit can be performed exactly and we get

\begin{equation}
\label{smart_metric_limit}
g(r; R \rightarrow 0) = 1 - \frac{16 \pi \rho_0}{3r} \left( r^3 - \frac{1}{\lambda} \ln{\left( \frac{1 + e^{\lambda r^3}}{2} \right)} \right) \ ,
\end{equation}

which still represents a RBH: indeed, in the asymptotic regimes, we have

\begin{equation}
\label{limit_limit_1}
g(r \rightarrow \infty; R \rightarrow 0) = 1 - \frac{2 M_0}{r} + o(r^{-2}) \ ,
\end{equation}

\begin{equation}
\label{limit_limit_2}
g(r \rightarrow 0; R \rightarrow 0) = 1 - \frac{8 \pi \rho_0}{3} r^2 - \frac{2 \pi \rho_0 }{3} \lambda r^5 + o(r^8) \ ,
\end{equation}

where $M_0 \equiv \frac{4}{3} \pi \left( (\ln 2 / \lambda)^{1/3} \right)^3 (2 \rho_0)$. Notice that eq. (\ref{limit_limit_2}) shows that the limits $\lambda \rightarrow \infty$ and $r \rightarrow 0$ cannot be inverted. This means that, in order to recover the classical limit, we should use eq. (\ref{smart_metric_limit}), which gives

\begin{equation}
\label{limit_limit_3}
g(r; R \rightarrow 0; \lambda \rightarrow \infty) = 1 - \frac{2M_0}{r} + o(\lambda^{-2}) \ .
\end{equation}

Since $M_0$ is proportional to $1/\lambda$, in the true classical limit it vanishes: i.e. metric (\ref{smart_metric_limit}) does not reduce to Schwarzschild, but already to Minkowski. Actually, this is not surprising, since the classical limit, in the regime of $ R \rightarrow 0$, is a mass distribution of null radius and finite constant density: 

\begin{equation}
\label{density_limit}
\rho (r; R=0; \lambda \rightarrow \infty) = \begin{cases} \rho_0, & r=0 \\ 0, & r \neq 0 \end{cases} \ .
\end{equation}

Finally, eq. (\ref{smart_metric_limit}) still represents a black hole: for both $r \rightarrow 0, \infty$ $g(r)$ is positive, while its minimum may still be negative; indeed, using the condition $g'(r_{min}) = 0$, we have

\begin{equation}
\label{minimum_limit}
g(r_{min}; R=0) = 1 - \frac{16 \pi \rho_0 r^2_{min}}{1 + e^{\lambda r^3_{min}}} \ ,
\end{equation}

which is either positive or negative, depending on the value of $\rho_0$.

\section{Conclusion}

In the present work we discussed how to build a RBH without working within the framework of GR, only using some exotic fluid instead of standard matter or vacuum. We showed that the standard maximal symmetry assumption $P = P_{\bot}$ is not able to produce not only a RBH, but also a singular one (at least in the SSSS framework): indeed there is no way to avoid or to circumnavigate the Buchdahl limit, independently on the specific choice of the density profile.

On the other hand, the choice $P = - \rho$ seems able to cure all these problems. We found that in this case, under an appropriate choice of the density profile, there is no Buchdahl limit to deal with and the classical radius $R$ of the star can be set small at will. In particular, it is interesting to note that $P = - \rho$ is the equation of state for dark energy, so that our model describes a dark energy black hole. It is questionable if such an object is realistic today, but it's credible that such black holes formed at the early stage of the universe, when dark energy was dominant.

It is worth to mention that such a black hole is really regular: indeed we still have to deal with the Schwarzschild term $c/r$, but, as we showed, there is a physical argument to set $c=0$, thus avoiding the central singularity. Since this is a general result, valid for any $R$, it holds also in the limit $R \rightarrow 0$.

A final observation is that we worked in a static framework, but in more points we introduced, though qualitatively, dynamical aspects (e.g. when we discussed the collapsing of the originating star to a black hole). Indeed, the static framework seems to be just an approximation of the full picture, which is expected to be time-dependent. On this light, our solution (\ref{smart_metric}) is expected to be only the static limit of some more complicated metric.

The investigation of the time-dependent solution, which is expected to be a more difficult effort, is also motivated by another point: as we said, it is possible that, during the early stage of the universe, a dark energy black hole could form. Since we proved that there is a residual black hole also when physical matter is no more present (see Sec. 7), we expect that the asymptotic behaviour of the time-dependent solution leaves a residual black hole (with $R=0$), which still continues to possess a mass: and these objects may be a candidate for present day dark matter. However, actually we don't have the whole dynamical analysis, so this is just a guess.

\bigskip

{\bf Acknowledgements}. We want to thank prof. L. Vanzo and prof. S. Zerbini for their fruitful assistance and discussions.


\begin{thebibliography}{9}

\bibitem{Schwarzschild}
	K. Schwarzschild,
	Sitzungsberichte Preuss. Akad. Wiss. Berlin (Math. Phys.), 424 (1916);
	English transl. by S. Antoci, [physics/9912033].

\bibitem{Reissner}
	H. Reissner,
	{\it Über die Eigengravitation des elektrischen Feldes nach der Einsteinschen Theorie},
	Ann. Phys. {\bf 50}, 106-120 (1916); doi: 10.1002/andp.19163550905.

\bibitem{Nordstrom}
	G. Nordström,
	{\it On the energy of the gravitational field in Einstein's theory},
	Verhandl. Koninkl. Ned. Akad. Wetenshap., Afdel. Natuurk., Amsterdam {\bf 26}, 1201-1208 (1918).

\bibitem{LIGO}
	 B.P. Abbott {\it et al.} [LIGO Scientific and Virgo Collaborations],
	{\it Observation of Gravitational Waves from a Binary Black Hole Merger},
	Phys. Rev. Lett. {\bf 116}, 061102 (2016).

\bibitem{Duan}
	Y.S. Duan,
	{\it Generalization of regular solutions of Einstein's gravity equations and Maxwell's equations for point-like charge},
	Soviet Physics JETP, {\bf 27} (6), 756-758 (1954);
	english. transl. by N. Korchagin, [gr-qc/1705.07752].

\bibitem{Bardeen}
	J.M. Bardeen,
	in Conference Proceedings of GR5 (Tbilisi, URSS, 1968), p. 174.

\bibitem{Hayward_1}
	S. Hayward,
	{\it Formation and evaporation of non singular black holes},
	Phys. Rev. Lett. {\bf 96}, 031103 (2006); doi: 10.1103/PhysRevLett.96.031103 [gr-qc/0506126].

\bibitem{Bronnikov}
	K.A. Bronnikov,
	{\it Regular magnetic black holes and monopoles from nonlinear electrodynamics},
	Phys. Rev. {\bf D 63}, 044005 (2001); doi: 10.1103/PhysRevD.63.044005 [gr-qc/0006014].

\bibitem{Elizalde}
	E. Elizalde, S.R. Hildebrandt
	{\it Family of regular interiors for nonrotating black holes with $T_0^0 = T_1^1$},
	Phys. Rev. {\bf D 65}, 124024 (2002); doi: 10.1103/PhysRevD.65.124024 [gr-qc/0202102v2].

\bibitem{Dymnikova}
	I. Dymnikova,
	{\it Regular electrically charged vacuum structures with de Sitter centre in nonlinear electrodynamics coupled to general relativity},
	Class. Quantum Grav. {\bf 21}, 4417 (2004); doi: 10.1088/0264-9381/21/18/009 [gr-qc/0407072].

\bibitem{Dymnikova92}
	I.Dymnikova,
	{\it Vacuum nonsingular black hole},
	Gen.Rel.Grav. {\bf 24}, 235 (1992).

\bibitem{NSS}
	P. Nicolini, A. Smailagic, E. Spallucci,
	{\it Noncommutative geometry inspired Schwarzschild black hole},
	Phys. Lett. {\bf B 632}, 547 (2006); doi: 10.1016/j.physletb.2005.11.004 [gr-qc/0510512].

\bibitem{ANSS}
	S. Ansoldi, P. Nicolini, A. Smailagic, E. Spallucci,
	{\it Noncommutative geometry inspired charged black holes},
	Phys. Lett. {\bf B 645}, 261 (2007); doi: 10.1016/j.physletb.2006.12.020 [gr-qc/0612035v1].

\bibitem{Modesto}
	S. Hossenfelder, L. Modesto, I. Prémont-Schwarz,
	{\it A model for non-singular black hole collapse and evaporation},
	Phys. Rev. {\bf D 81}, 044036 (2010); doi: 10.1103/PhysRevD.81.044036 [gr-qc/0912.1823v3].

\bibitem{CMSZ}
	G. Cognola, R. Myrzakulov, L. Sebastiani, S. Zerbini,
	{\it Einstein gravity with Gauss-Bonnet entropic corrections},
	Phys. Rew. {\bf D 88}, 024006 (2013); doi: 10.1103/PhysRevD.88.024006 [gr-qc/1304.1878v2].

\bibitem{Dymnikova_2}
	I. Dymnikova, E. Galaktionov,
	{\it Regular rotating electrically charged black holes and solitons in nonlinear electrodynamics minimally coupled to gravity},
	Class. Quantum Grav. {\bf 32}, 165015 (2015); doi: 10.1088/0264-9381/32/16/165015 [gr-qc/1510.01353v1].

\bibitem{Culetu}
	H. Culetu,
	{\it Microscopic corrections to Schwarzschild spacetime}, (2015) [gr-qc/1508.07570v2].

\bibitem{Maeda}
	G. Kunstatter, H. Maeda, T. Taves,
	{\it New two-dimensional effective actions for non-singular black holes}, (2015) [gr-qc/1509.06746v2].

\bibitem{Horava}
	P. Horava,
	{\it Membranes at quantum criticality},
	JHEP {\bf 0903} 20 (2009); doi: 10.1088/11

\bibitem{Horava_1}
	P. Horava,
	{\it Quantum gravity at a Lifshitz point},
	Phys. Rev. {\bf D 79} 084008 (2009); doi: 10.1103/PhysRevD.79.084008 [hep-th/0901.3775v2].

\bibitem{KS}
	A. Kehagias, K. Sfetsos,
	{\it The black hole and FRW geometries of non-relativistic gravity},
	Phys. Lett. {\bf B 678}, 123 (2009); doi: 10.1016/j.physletb.2009.06.019 [hep-th/0905.0477v1].

\bibitem{CCO}
	R.G. Cai, L.M. Cao, N. Ohta,
	{\it Black holes in gravity with conformal anomaly and logarithmic term in black hole entropy},
	JHEP {\bf 1004}, 082 (2010); doi: 10.1007/JHEP04(2010)082 [hep-th/0911.4379v2].

\bibitem{Pradhan}
	P. Pradhan,
	{\it Area (or entropy) product formula for a regular black hole}, (2015) [gr-qc/1512.06187].

\bibitem{Ma}
	M.S. Ma,
	{\it Magnetically charged regular black hole in a model of nonlinear electrodynamics},
	Annals Phys. {\bf 362}, 529 (2015); doi:  10.1016/j.aop.2015.08.028 [gr-qc/1509.05580].

\bibitem{Johannsen}
	T. Johannsen,
	{\it Regular black hole metric with three constants of motion},
	Phys. Rev. {\bf D 88}, 044002 (2013); doi: 10.1103/PhysRevD.88.044002 [gr-qc/1501.02809v2].

\bibitem{Rodrigues}
	M.E. Rodrigues, J.C. Fabris, E.L.B. Junior, G.T. Marques
	{\it Generalization of regular black holes in General Relativity to $f(R)$ gravity},
	EPJ {\bf C 76}, 250; doi: 10.1140/epjc/s10052-016-4085-x [gr-qc/1601.00471].

\bibitem{Fan}
	Z.Y. Fan and X. Wang,
	{\it Construction of Regular Black Holes in General Relativit}, arXiv:1610.02636 [gr-qc].

\bibitem{Beato}
	E. Ayon–Beato, A. Garcia,
	{\it Regular black hole in general relativity coupled to nonlinear electrodynamics},
	Phys. Rew. Lett. {\bf 80}, 5056 (1998); doi: 10.1103/PhysRevLett.80.5056 [gr-qc/9911046v1].

\bibitem{Frolov_1}
	V.P. Frolov, A. Zelnikov,
	{\it Quantum radiation from a sandwich black hole},
	Phys. Rev. {\bf D 95}, no. 4, 044042 (2017); doi:10.1103/PhysRevD.95.044042 [arXiv:1612.05319 [hep-th]].

\bibitem{Frolov_2}
	V.P. Frolov, A. Zelnikov,
	{\it Quantum radiation from an evaporating non-singular black hole}, [arXiv:1704.03043 [hep-th]].

\bibitem{Frolov_3}
	V.P. Frolov,
	{\it Notes on non-singular models of black holes},
	Phys. Rev. {\bf D 94}, no. 10, 104056 (2016); doi:10.1103/PhysRevD.94.104056 [arXiv:1609.01758 [gr-qc]].

\bibitem{Frolov_4}
	V.P. Frolov,
	{\it Information loss problem and a ’black hole‘ model with a closed apparent horizon},
	JHEP {\bf 1405}, 049 (2014); doi:10.1007/JHEP05(2014)049 [arXiv:1402.5446 [hep-th]].

\bibitem{Frolov_5}
	V.P. Frolov, M.A. Markov, V.F. Mukhanov,
	{\it Through A Black Hole Into A New Universe?},
	Phys. Lett. {\bf B 216}, 272 (1989); doi:10.1016/0370-2693(89)91114-3;

\bibitem{Frolov_6}
	V.P. Frolov, M.A. Markov, V.F. Mukhanov,
	{\it Black Holes as Possible Sources of Closed and Semiclosed Worlds},
	Phys. Rev. {\bf D 41}, 383 (1990); doi:10.1103/PhysRevD.41.383.

\bibitem{Frolov_7}
	P.A. Bolashenko, V.P. Frolov,
	{\it Certain Properties Of A Nonsingular Model Of A Black Hole}
	in *MARKOV, M.A. (ED.): THE PHYSICAL EFFECTS IN THE GRAVITATIONAL FIELD OF BLACK HOLES* 205-218.

\bibitem{ans} 
	S. Ansoldi,
	{\it Spherical black holes with regular center: A Review of existing models including a recent realization with Gaussian sources}, arXiv:0802.0330 [gr-qc].

\bibitem{Sakharov} 
	A. Sakharov,
	{\it Initial stage of an expanding universe and appearance of a nonuniform distribution of matter},
	Sov. Phys. JETP {\bf 22}, 241 (1966).

\bibitem{BLZ}
	A.B. Balakin, J.P.S. Lemos and A.E. Zayats,
	{\it Regular nonminimal magnetic black holes in spacetimes with a cosmological constant},
	Phys.Rev.D {\bf 93}, no. 2, 024008 (2016); doi:10.1103/PhysRevD.93.024008 [arXiv:1512.02653 [gr-qc]].

\bibitem{Sert}
	Ö. Sert,
	{\it Regular black hole solutions of the non-minimally coupled $Y(R)F^2$ gravity},
	Journal of Math. Phys. {\bf 57}, 032501 (2016); doi: 10.1063/1.4944428 [gr-qc/1512.01172v2].

\bibitem{Ovalle}
	J. Ovalle,
	{\it Decoupling gravitational sources in general relativity: from perfect to anisotropic fluids}, [gr-qc/1512.01172v2].

\bibitem{Babichev_1}
	E. Babichev, V. Dokuchaev, Y. Eroshenko,
	{\it Black hole mass decreasing due to phantom energy accretion},
	Phys. Rev. Lett. {\bf 93}, 021102 (2004); doi: 10.1103/PhysRevLett.93.021102 [gr-qc/0402089v3].

\bibitem{Babichev_2}
	E. Babichev, V. Dokuchaev, Y. Eroshenko,
	{\it The accretion of dark energy onto a dlack hole},
	J. Exp. Theor. Phys. {\bf 100}, 528-538 (2005); doi: 10.1134/1.1901765 [gr-qc/0505618v1].

\bibitem{Babichev_3}
	E. Babichev, V. Dokuchaev, Y. Eroshenko,
	{\it Black holes in the presence of dark energy},
	Phys. Usp. {\bf 56}, 1155-1175 (2013); doi: 	10.3367/UFNe.0183.201312a.1257 [gr-qc/1406.0841v1].

\bibitem{Mazur_Mottola}
	P.O. Mazur, E. Mottola,
	{\it Surface Tension and Negative Pressure Interior of a Non-Singular 'Black Hole'},
	Class. Quantum Grav. {\bf 32}, 21 (2015); doi: 10.1088/0264-9381/32/21/215024 [gr-qc/1501.03806v1].

\bibitem{Nicolini}
	P. Nicolini, A. Smailagic, E. Spallucci,
	{\it Reply to arXiv:1704.08516 "A note on singular and non-singular black holes"}, [gr-qc/1705.05359].	

\bibitem{CZ}
	S. Chinaglia, S. Zerbini,
	{A note on singular and non-singular black holes},
	Gen. Reltiv. Gravit. {\bf 49}, 75 (2017); doi: 10.1007/s10714-017-2235-6 [gr-qc/1704.08516].

\bibitem{Buchdahl}
	H.A. Buchdahl,
	{\it General Relativistic Fluid Spheres},
	Phys. Rev. {\bf 116} (1959) 1027.

\end{thebibliography}
\end{document}